\documentclass[aps,prl,reprint,floatfix,amsmath,amssymb,superscriptaddress]{revtex4-1}

\usepackage{graphicx}
\usepackage{epstopdf}
\usepackage{color}
\usepackage{hyperref}
\usepackage{bm}
\usepackage{printlen}
\usepackage{units}

\usepackage[table]{xcolor}

\renewcommand{\vec}[1]{\mathbf{#1}}

\renewcommand{\L}{\mathcal{L}}
\newcommand{\G}{\mathcal{G}}

\hypersetup{hidelinks}

\begin{document}

\title{Quantum transport  in dispersionless electronic bands}

\author{Alexander  Kruchkov}   

\affiliation{Institute of Physics, Swiss Federal Institute of Technology (EPFL),  Lausanne, CH 1015, Switzerland}

  \affiliation{Department of Physics, Harvard University, Cambridge, Massachusetts 02138, USA} 

\affiliation{Branco Weiss Society in Science, ETH Zurich, Zurich, CH 8092, Switzerland} 


\begin{abstract}
Flat electronic bands are counterintuitive: with the electron velocity vanishing, our conventional notions of quasiparticle transport are no longer valid. We here study the quantum transport in the generalized families of perfectly flat bands [PRB 105, L241102 (2022)], and find that while the conventional  contributions indeed vanish, the quantum-geometric contribution gives rise to the enhanced electronic transport. This contribution is connected to the Wannier orbital quantization in the perfectly flat bands, and is present only for geometrically-nontrivial bands (for example, flat Chern bands).  We find structurally similar expressions for thermal conductance, thermoelectric response, and superfluid weight in the flat bands. In particular, we report the anomalous thermopower associated with flat topological bands reaching values as large as $\frac{k_B}{e}\ln 2$$\approx$60 $\mu$V/k, the quantum unit of thermopower, which are not expected for the  conventional dispersive bands.  
\end{abstract}

\maketitle


Dispersionless electronic states ("flat bands") are counterintuitive since the effective electronic mass becomes infinite, the quasiparticle velocity vanishes, and the conventional  notions of electron transport fail.  
The perfectly flat electronic bands is a change of paradigm in condensed matter physics, but  remained largely a hypothetical object until the discovery of the twisted bilayer graphene \cite{Cao2018a,Cao2018b}, where the dispersionless electronic states are  emerging at the magic angle $1.05^{\circ}$\cite{Trambly2010, Bistritzer2011}. 
The underlying flat band is not just a lucky engineering of material parameters, but it is of fundamental origin as it can be tuned to perfectly flat \cite{TKV2019,SanJose2012}. It was further understood that the magic-angle flat bands are dual to the lowest Landau level, and host a number of unconventional phases, characterized by strange metallicity \cite{Polshyn2019,Cao2020,Jaoui2022}, unconventional superconductivity \cite{Cao2018b,Lu2019}, fractional Hall conductance \cite{Xie2021}, and giant thermopower \cite{Paul2022}---untypical for conventional electronic systems.

A recent interest in condensed matter physics is understanding the role quantum geometry of electronic states, described by the quantum geometric tensor \cite{Provost1980, Ma2010}
\begin{align}
\mathfrak G^{(n)}_{ij} = \langle \partial_{k_i} u_{n \vec k} | \left [1-      | u_{n \vec k} \rangle \langle u_{n \vec k}  |  \right] | \partial_{k_j} u_{n \vec k} \rangle
\label{metric}
\end{align}
here $u_{n \vec k}$ is the associated Bloch state of the $n$th electronic band, which may or may not be flat. 
The real part of $\mathcal G_{ij}$$=$$\text{Re} \mathfrak G_{ij} $,  
is the Fubini-Study metrics  describing the geometry of the bands, while the imaginary part $ \mathcal F_{ij}$$=$$- 2 \text{Im} \mathfrak G_{ij}$ is  Berry curvature reflecting the topology of the Bloch states.  In particular, it has been understood that the largely overlooked quantum metrics plays important role in different quantum transport phenomena, ranging from quantum noise, optical conductivity, anomalous Hall effect, and unconventional superconductivity, and adjacent topics \cite{Neupert2011, Haldane2004, Rhim2020, Peotta2015, Huhtinen2022,  Gao2019, Lapa2019, Kozii2021, Mitscherling2022, Ahn2022, Chen2021}. The quantum-geometric superconductivity \cite{Peotta2015} has a particular important role in twisted bilayer graphene, where it has been shown that the quantum-geometric contribution to the superfluid weight is key at the magic angle \cite{Hu2019,Hazra2019,Julku2020,Xie2020}. Moreover, it has recently been argued that the quantum-geometric contribution to superfluid weight it TBG can be probed in experiment with the ultraclean samples \cite{Tian2021}. However, other anomalies in the topological flat bands,---such as e.g. giant thermoelectric power at the magic angle \cite{Paul2022}---remain to be revisited from the quantum geometric perspective as well.

\begin{figure}[t]
\includegraphics[width = 0.9 \columnwidth]{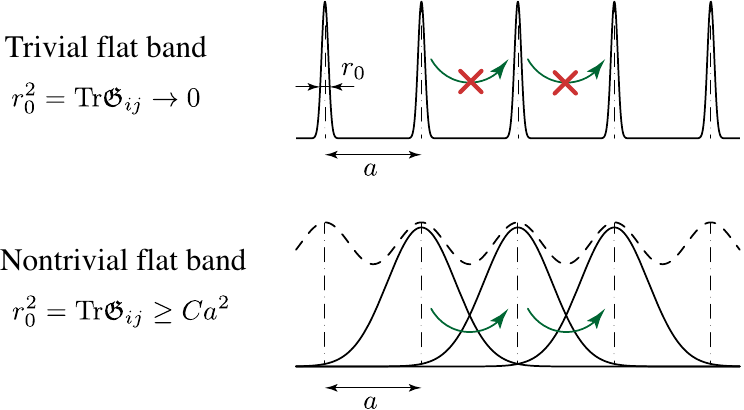}
\caption{(Top) In trivial flat bands, electrons are strongly localized (Wannier orbitals sharp) and  electronic transport is forbidden, the system is in the insulating phase. (Bottom) In nontrivial flat bands (e.g. Chern bands), the Wannier orbitals cannot be exponentially localized; electrons do not move in the classical sense.  Under application of external fields, electrons tunnel between overlapping Wannier orbitals, resulting into  unconventional conductivity without  electron velocities. }
\label{Fig1}
\end{figure}

In this direction, recent classification of perfectly flat bands \cite{Kruchkov2022} presents a handy framework as it allows to bridge the properties of Wannier orbitals, their quantum geometry, and band flatness limits. 
 If we allow perfect band flatness, the quantum metrics  of dispersionless electronic bands saturates the "trace condition" 
\begin{align}
\text{Tr} \, \mathcal G_{ij} (\vec k)  = |\mathcal F_{xy} (\vec k)|. 
\label{trace}
\end{align}
However, the nontrivial quantum geometry comes at cost that the electronic Wannier orbitals will have a finite and large cross-section \cite{Marzari1997}.  In case of perfectly flat Chern bands \eqref{trace}, the Wannier orbital crossection $r_0^2$ experiences Lifshitz-Onsager-like  quantization \cite{Kruchkov2022}, with
\begin{align}
r_0^2 = a^2 \int d \vec k \text{Tr} \mathcal G_{ij} (\vec k)  =  a^2 \int d^2 \vec k  \mathcal F_{xy} (\vec k)  = C a^2. 
\end{align}
 In other words, the dispersionless electronic states are  spread over the atomic lattice, overlapping with $C$ neighboring electronic orbitals, hence allowing quantum tunneling even in the absence of kinetic terms
(Fig.\ref{Fig1}). We show that this phenomenon promotes the unconventional quantum transport in dispersionless electronic bands.

In this work we bring to the common denominator different quantum transport properties (electric conductivity, quantized Hall effect,  thermal conductivity, thermoelectric response, and  the superfluid weight), which for the perfectly flat band systems can be expressed through their multiorbital quantum metrics: 
\begin{align}
\mathfrak {L}_{ij} =   \sum_{nm, \vec k} \mathcal I_{nm} (\vec k) \, \text{Re} \mathfrak G^{nm}_{ij}  (\vec k) 
+ \mathcal J_{nm} (\vec k) \, \text{Im} \mathfrak G^{nm}_{ij}  (\vec k), 
\nonumber
\end{align}
where $\mathfrak {L}_{ij}$ is a quantum transport characteristic,  and $\mathfrak G^{nm}_{ij}  (\vec k) \equiv \langle \partial_{k_i} u_{n\vec k} | u_{m\vec k}  \rangle \langle u_{m\vec k} |  \partial_{k_j} u_{n\vec k}  \rangle$ is the generalized quantum-geometric tensor (defined further in text), and $\mathcal I_{nm} (\vec k)$ and $\mathcal J_{nm} (\vec k)$ are system-dependent structural tensors expressed through quasiparticle propagators. Note that neither the quasiparticle velocities nor bandwidth enter this expression; this flat-band transport is purely quantum, with its origin in Wannier functions overlap (Fig.\ref{Fig1}).

\textbf{Quantum transport formalism.} In what follows below we consider a weakly-dispersive Chern band, and then set the bandwidth (and hence the Fermi velocity) to \textit{exact zero}. The main result is illustrated for the perfectly flat Chern bands, however it also applies to all geometrically-nontrivial  flat bands (Wannier function are not exponentially localized), thus including those  in twisted bilayer graphene and similar materials. For the moment, we omit the explicit dependence on  magnetic fields, however such generalization can be  done.  
Up to this moment, the derivation of polarization tensor is rather conventional and can be obtained in several ways. A disciplined way to derive it is through using the current-current correlators in Matsubara framework \cite{Mahan}.  
The quantum transport properties  are computed through imaginary-time Matsubara correlators
\begin{align}
\L^{\alpha \beta}_{ij} (\tau ,\tau') = - \frac{1}{\hbar} 
\langle \mathcal{T}_{\tau} \, J^{\alpha}_i (\tau) Q^{\beta}_j(\tau')  \rangle ,
\label{Onsager-1}
\end{align}
where $Q^{\alpha}$ is the generalized "charge" operator (we denote $\alpha= 1$  for electric charge, $\alpha'=2$ for heat transfer), $i,j = x,y$ and $\vec J^{\alpha}$ is the generalized current. 
For example, for the electric current of charge $e$ one writes \cite{Mahan} 
\begin{align}
\vec J = \frac{e}{\hbar} \sum_{\vec k}  c^{\dag}_{\vec k} \frac{\partial \mathcal H_{\vec k}}{\partial \vec k}  c^{}_{\vec k} .
\label{current}
\end{align}
To calculate the response functions $\L_{ij}$, we introduced the auxiliary current-current correlators
\begin{align}
\Pi^{\alpha \beta}_{ij} (\tau) = -  
\langle \mathcal T_{\tau} \,  J^{\alpha}_i (\tau ) J^{\beta}_j(0)  \rangle,
\label{current-current}
\end{align}
so in the frequency representation one has 
$
\L_{ij}(i \omega_n) = \frac{1} {i \omega_n} \left[ \Pi_{ij} (i \omega_n)-\Pi_{ij} (0)\right] .
$
To calculate the transport properties (such as conductance $\sigma_{ij} = \L^{11}_{ij} $), we further proceed to analytical continuation of $T_{ij} (i \omega_n)$ and then take the DC limit:
\begin{align}
\L^{\alpha \beta}_{ij} \equiv  \lim_{\omega \to 0}  \frac{\Pi^{\alpha \beta}_{ij} ( \omega)  - \Pi^{\alpha \beta}_{ij} (0) } {i \omega} 
\label{Onsager} .
\end{align}
The Onsager coefficients $\L^{\alpha \beta}_{ij}$ fully describe the transport properties of an electronic system. 
 Experimentally, the transport measurements across the sample are performed by $I_e = \L_{11} \Delta V + \L_{12} \Delta T$ (electric measurement)  and $I_h =  \L_{21} \Delta V + \L_{22} \Delta T $ (heat measurement) \cite{OnsagerA,*OnsagerB, Rammer2018}. We further compute the electric conductivity $\sigma_{ij} =\L_{11} $, thermal conductivity $\kappa_{ij} = \beta(\L_{22} - L^2_{12}/T \L_{11})$, and thermoelectric response (Seebeck coefficient $\Theta = \frac{\beta} \L_{12}/\L_{11}$) in the dispersionless electronic bands (here $\beta$$=$$1/T$; $e$ is included in definition of Eq.\eqref{current}). A similar response structure to  vector potential $\vec A$  will  imply the finite superfluid weight $D_{\text{S}}$ in the dispersionless bands \cite{Huhtinen2022}.

We start from the electric conductivity, for which we calculate the electric polarization tensor $\Pi_{ij}(\omega)$ (we here drop the superscripts $\alpha \beta$ for brevity); for convenience, below we use    $\Pi_{ij}(\omega) = \frac{e^2}{\hbar^2} \tilde \Pi_{ij}(\omega)$.  Evaluating the current-current correlator  \eqref{current-current} with \eqref{current} in Matsubara representation gives \footnote{We introduce here two quantities $\tilde \Pi^{\pm} _{ij}$ in order to evaluate carefully zero-frequency limit in Eq.\eqref{Onsager}} 
 \begin{align}
\tilde \Pi^{\pm} _{ij} (i \omega_0)   =  \frac{1}{ \beta } \sum_{\vec k} \sum_{i \omega'_n} \text{Tr} \, 
 \G_{\vec k} ( i \omega'_n) \frac{\partial \mathcal H_{\vec k}}{\partial k_i} 
 \G_{\vec k} (i \omega'_n \pm i \omega_0) \frac{\partial \mathcal H_{\vec k}}{\partial k_j}  ,
 \nonumber 
 \label{polarization}
\end{align}
where  $\G_{\vec k} (i \omega')$ is the Matsubara transform of the (renormalized) Green function $\G_{\vec k} (\tau, \tau') = - \langle \mathcal T_{\tau} \, 
c^{\dag}_{\vec k} (\tau) c^{}){\vec k} (\tau')  \rangle$, where expectation value is taken over the interacting vaccum at temperature $T$. Here Matsubara frequencies $i \omega'_n$  are fermionic and $i \omega_0$ is bosonic.

The influence of the quantum-geometric tensor can be demonstrated in the following way. The position operator in the Bloch basis is \cite{Blount1962}
\begin{align}
\hat{\vec r}_{mn} = i \partial_{\vec k} \delta_{nm} +  
\langle u_{n \vec k} | \, i \partial_{\vec k}  \,
u_{m\vec k} \rangle .
\end{align}
It follows that  the generalized velocity operator in this basis is given by
\begin{align}
\frac{1}{\hbar} \frac{\partial \mathcal H_{\vec k}}{\partial \vec k}  = \dot{\vec r}  = \vec v_{n \vec k} \delta_{nm}  + \omega_{nm,\vec k} \, \langle u_{n \vec k} | \partial_{\vec k}  
u_{m\vec k} \rangle  , 
\label{velocity}
\end{align}
where $\vec v_{n \vec k} = \frac{\partial \varepsilon_{n \vec k}}{ \partial \hbar {\vec k} } $ is the quasiparticle velocity in the band ("Fermi velocity") and $\hbar \omega_{nm,\vec k} = \varepsilon_{n \vec k} - \varepsilon_{m \vec k} $ are transition frequencies of the  multiorbital system  (the second term is  called "anomalous velocity" \cite{Haldane2004}).
 With the velocity operator \eqref{velocity}, the polarization tensor \eqref{polarization} have two terms: the first proportional to the Fermi velocities $\mathcal O (\vec v_{n \vec k}^2)$, and the second being independent of the band dispersion itself. For illustrational purpose, it is useful to write down the first  contribution which has a generic form for the $n$-th band
$
\sigma_0 = \frac{e^2}{ \hbar}  \sum_n \sum_{\vec k}   \frac{\partial \varepsilon_{n \vec k}}{\partial k_i}  
\frac{\partial \varepsilon_{n \vec k}}{\partial k_i}  S_n (\vec k)  = 0 
$.
The conventional  longitudinal contribution vanishes exactly since  the electronic band is perfectly flat (witnessed in zero Fermi velocity $\partial_{\vec k} \varepsilon_{n \vec k}$$\equiv$$0$ in all the Brillouin zone).  This is where our conventional intuition  comes to an edge.

In contrast, by using Eqs.\eqref{Onsager}-\eqref{velocity}, we  find that there is 
a  quantum-geometric contribution to the DC transport even in the case of perfectly flat bands,
\begin{align}
\sigma_{ij} =  \frac{e^2}{\hbar}  \sum_{\vec k} \sum_{nm}  & \mathcal I_{nm} (\vec k) \, \text{Re}  \mathfrak G^{nm}_{ij}  (\vec k) 
\nonumber
\\
+   & \mathcal J_{nm} (\vec k) \, \text{Im} \mathfrak G^{nm}_{ij}  (\vec k), 
\label{flat-sigma}
\end{align}
where we have introduced the generalized geometric tensor for multiorbital system
\begin{align}
\mathfrak G^{nm}_{ij}  (\vec k) \equiv \langle \partial_{k_i} u_{n\vec k} | u_{m\vec k}  \rangle \langle u_{m\vec k} |  \partial_{k_j} u_{n\vec k}   \rangle ;
\end{align}
note that $\sum_{m\ne n} \mathfrak G^{nm}_{ij}  = \mathfrak G^{(n)}_{ij}$, with $\mathfrak G^{(n)}_{ij}$ given by formula \eqref{metric}. 
The structural tensors  $\mathcal I_{nm} (\vec k)$ and $\mathcal J_{nm} (\vec k)$ are fully determined by the quasiparticle propagators; they are defined in the symmetrized form as (see SM \cite{SM2022})
\begin{align}
\mathcal I_{nm}(\vec k) & = - 2 \omega^2_{nm,\vec k} \int \limits_{-\infty}^{+\infty}  \frac{ d \omega} { 2 \pi }   \,  f'(\omega)  \,     
\text{Im} \G^{R}_{n \vec k} (\omega) \text{Im} \G^R_{ m \vec k} (\omega) , \nonumber
\\
\mathcal J_{nm}(\vec k) & =  8  \, \omega^2_{nm,\vec k}   \iint \limits_{-\infty}^{+\infty}  
\frac{ d \omega  d \omega' }{(2 \pi)^2}   \, 
 \frac{ 
 f(\omega) \,   
\text{Im} \G^R_{n \vec k} (\omega') \text{Im} \G^R_{ m \vec k} (\omega) 
}
{
(\omega - \omega')^2
}
,  \nonumber
\end{align}
with $\omega_{nm,\vec k}  = \varepsilon_{n \vec k}  - \varepsilon_{m \vec k}$, and $f(w)$ is the Fermi-Dirac distribution function.  
Generically, the structural tensors $\mathcal I_{nm}$  and $\mathcal J_{nm}$ are  non-zero, and depend on the nature of quasiparticles in the system, described by causal (retarded) propagators $\G^R_{ m \vec k} (\omega)$. 
 Formula \eqref{flat-sigma} is working for the many-body propagators with well-defined quasiparticle poles; the quasiparticle energy bands $\varepsilon_{n \vec k}$ may or may not have dispersion as such.

\textbf{Absence of semiclassical analogues.}
The first term in  Eq.\eqref{flat-sigma} is a  quantum effect which has no classical analogues; to our knowledge the   longitudinal conductance reminding Eq.\eqref{flat-sigma} was  obtained in the clean limit by Resta  \cite{Resta2005}.
The quasiparticle propagator  $Z_{\vec k}/(\omega - \Sigma (k, \omega) )$ in the clean limit is characterized by $\Sigma (\vec k, \omega)$$\to$$i \delta$, $\delta$$\to$$0$, hence Eq.\eqref{flat-sigma} gives $\sigma_{xx} (\omega$$=$$0)  \propto \delta$, which vanishes for small $\delta$  as expected (see Fig. 2b). From another perspective, the flat band quantum transport \eqref{flat-sigma} is  determined by the quantum metrics of the electronic bands $\mathfrak G^{nm}_{ij}  (\vec k)
$; since the Fubini-Study metrics is inherently related to the uncertainty principle \cite{Braunstein1994,Anandan1990},  the longitudinal  ($\sigma_{xx}$) term in \eqref{flat-sigma} has no classical analogues for the  "Drude" electrons. 
The transverse ($xy$) term in \eqref{flat-sigma} neither represents the classical case, as it give rise to the quantized Hall conductance, and is connected to TKNN invariant \cite{TKNN}. To illustrate this, we consider a simple model with flat Chern bands below.

\begin{figure*}[t]
\includegraphics[width = 0.7 \textwidth]{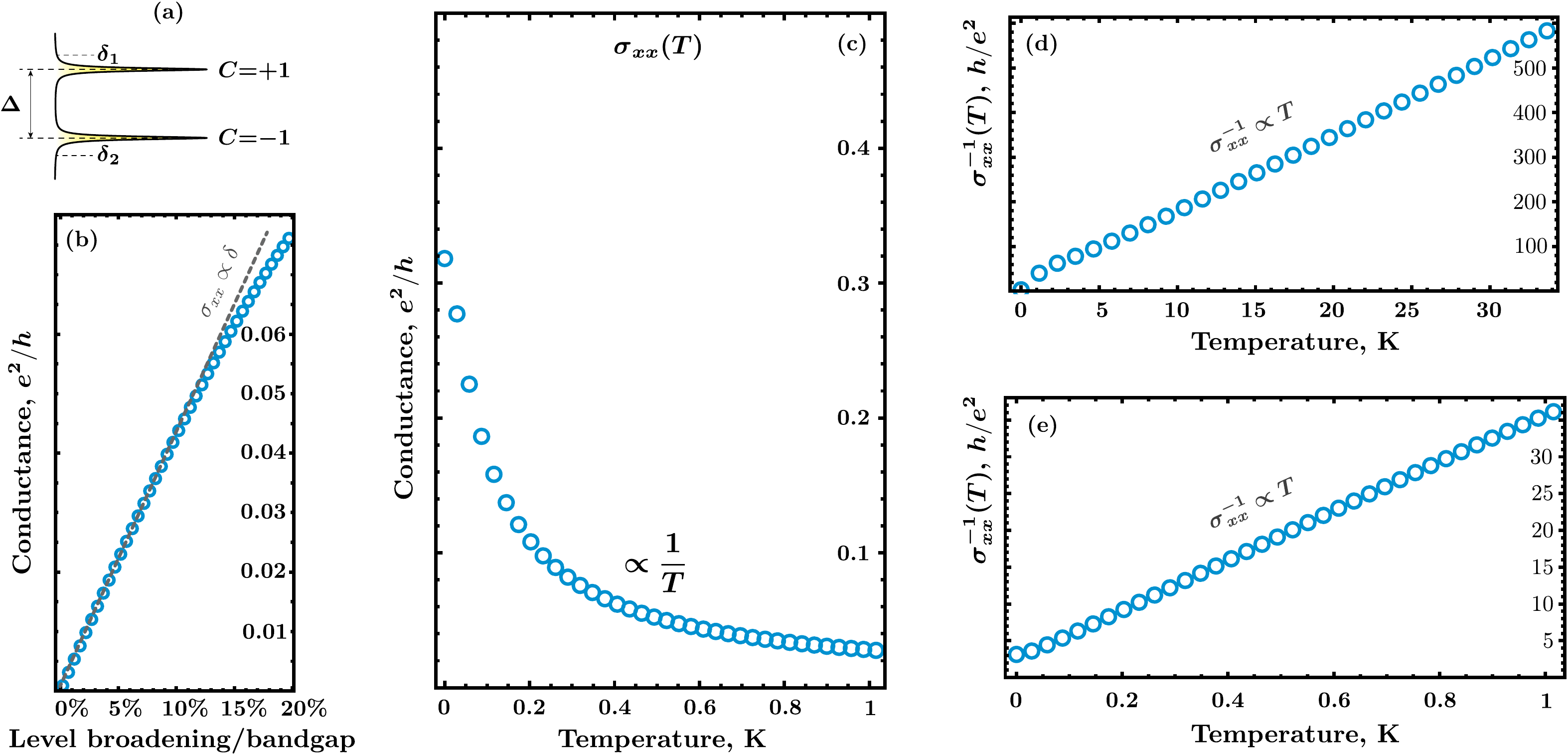}
\caption{\textbf{Longitudinal electronic transport in topological flat  bands.} (a) Schematic of a two-level system with two topological flat bands of $C= \pm 1$. (b) Longitudinal conductance as function of level broadening calculated here for the band gap $\Delta = 0.5$ meV and temperature $T$$=$$\Delta$ ($\approx$ 6 K); here $\delta_{1,2}$$\equiv$$\delta$. (c) Longitudinal conductance as function of temperature, calculated at $\Delta = 0.5$ meV, $\delta_{1,2}$$=$$0.01$ meV. (d,e) Inverse longitudinal conductance (for the same parameters) shows T-linear scaling both for high temperatures ($T$$ \gg$$\Delta$) and low temperatures ($T$$\ll$$\Delta$); $T$-slopes in different regimes  (d) and (e) are slightly different.  
}
\label{Fig2}
\end{figure*}

\textbf{Quantum Hall effect.}   In this key example we focus on the transverse response associated with the perfectly flat Chern bands.  The transverse (Hall) conductivity determined through quasiparticle propagators is given by  Eq. \eqref{flat-sigma} at  $T=0$  as \begin{align}
\sigma_{xy} =  \frac{ 8
 e^2}{ \hbar}  \sum_{\vec k}   \sum_{n  m}{}^{'}   
\iint \limits_{-\infty}^{+\infty} \frac{d \omega  d \omega'}{(2 \pi)^2}  & 
 \frac{ 
\,   
\text{Im} \G^R_{n \vec k} (\omega') \text{Im} \G^R_{ m \vec k} (\omega) 
}
{
(\omega - \omega')^2 
} 
\nonumber
\\
\times &
\omega^2_{nm,\vec k}    \, \text{Im} \mathfrak G^{nm}_{xy}  (\vec k) ; 
\label{Hall}
\end{align}
the summation index $n$ runs over the occupied bands (below Fermi level $\varepsilon_F$).
In the limit when the quasiparticles are well-defined, one may use $\text{Im} \G^R_{n \vec k} (\omega) 
 = \delta_{n \vec k} / [ (\omega - \omega_{n \vec k} + i \delta_{n \vec k} ) (\omega - \omega_{n \vec k} - i \delta_{n \vec k} ) ] $, the expression \eqref{Hall} contains integrands with pole singularities and we can resort to residue theorem to evaluate integrals for aribtrary $\delta_{n \vec k}$, $\delta_{m \vec k}$ (see SM \cite{SM2022}). 
By choosing an appropriate contour, we obtain $ \iint  _{-\infty}^{+\infty}  
 \frac{ 
\,   
\text{Im} \G^R_{n \vec k} (\omega') \text{Im} \G^R_{ m \vec k} (\omega) 
}
{
(\omega - \omega')^2 
} d \omega  d \omega'  \simeq  \pi^2 [{\omega_{n \vec k} - \omega_{m \vec k}}]^{-2} \left[1 + \mathcal{O} (\delta^2) \right] \approx  \pi^2 /\omega_{nm, \vec k}^2$. 
Using this expression for formula \eqref{Hall},  in the limit of well-resolved energy bands ($\delta_{n \vec k} , \delta_{m \vec k} \ll \Delta$, where $\Delta$ is the band gap), we obtain the established quantum Hall conductance 
\begin{align}
\sigma_{xy} = & \frac{e^2}{\hbar} \sum_{n} {}^{'} \sum_{\vec k}    \mathcal F^{(n)}_{xy}  (\vec k) = \frac{e^2}{h} \sum_n {}^{'} C_n. 
\end{align}
where $ \mathcal F^{(n)}_{xy}$ is the Berry curvature of the $n$-th band. Thus we obtain the quantum Hall conductance in the form of TKNN invariant \cite{TKNN}.   Note that the quantized nature of \eqref{Hall} holds  in the presence of moderate interactions, provided the renormalized propagators have well-defined quasiparticle poles.  Note also, that when the Fermi level is within the flat band, we obtain anomalous Hall  contributions similar to Refs. \cite{Haldane2004,Jungwirth2002}. Importantly, formula \eqref{Hall} provides means of calculation for quantum Hall response in perfectly flat bands in the case when the momentum dispersion of the effective Hamiltonian $\mathcal H (\vec k)$ is not defined and conventional minimal coupling to magnetic vector potential as in approach \cite{TKNN} is not applicable.

\textbf{Longitudinal conductance}. For storytelling, consider the Haldane model \cite{Haldane1988} on NN and NNN hoppings ($\Lambda =2$), 
\begin{align}
\mathcal H_0   = \sum_{i} t_0 c^{\dag}_i c^{}_i + \sum_{\langle ij \rangle} t^{\text{NN}}_{ij} c^{\dag}_{i } c^{\dag}_{j } +  \sum_{\langle\langle ij \rangle \rangle} t^{\text{NNN}}_{ij} c^{\dag}_{i } c^{\dag}_{j },
\label{Haldane}
\end{align}
where we fix $t^{\text{NN}}_{ij}$ as real and $t^{\text{NNN}}_{ij}$$=$$t'e^{i \Phi_{ij}}$. Clearly, for $\Phi_{ij}$$=$$\pm \pi/2$, the spectrum becomes particle-hole symmetric \cite{Haldane1988}. This yields upon transformation (1)  two pefectly flat bands positioned at $E$$=$$\pm E_0$ \cite{Kruchkov2022}. The resulting $\mathcal H_{0}^{\textbf{flat}}$ becomes nonlocal ($\Lambda'$$=$$\infty$), though it is possible to truncate it and make the band arbitrary flat by choosing a corresponding truncation $\Lambda'$ and minimizing bandwidth (Refs.\cite{Neupert2011,Kruchkov2022}). Hence by allowing further hopping terms to \eqref{Haldane}, this model features two dispersionless Chern bands with $C=\pm 1$. 
By using the flat band Green's function matrix  $1/(\omega- \mathcal H_{0}^{\textbf{flat}})$ and introducing  level broadening $i \delta$, we use Eq. \eqref{flat-sigma} to calculate $\sigma_{xx}$.  

The longitudinal conductance $\sigma_{xx}$ of the flat bands has an unconventional temperature dependence:
 We find that quite generically the  $\sigma_{xx}$ conductance has inverse temperature scaling (Fig. 2) 
\begin{align}
\sigma_{xx} (T)  \propto  \frac{\sum_{\vec k} \text{Tr} \, \mathcal G_{ij} (\vec k) } {T}  \sim \frac{1}{T} , 
\label{strange}
\end{align}
reminiscent of strange metal phases.  The quantity in nominator is   in order of $\mathcal O (1)$ (for a perfectly flat Chern band, it is proportional to $|C|$ of the flat band).  This $T^{-1}$ scaling is quite universal, and while the particular numbers depend on the values of $\delta$ (Fig. 2a), the $\sigma^{-1}_{xx}  \sim T$ scaling is seen both for high temperatures ($T$$\ll$$\Delta$)   and high temperatures ($T$$\gg$$\Delta$), (Figs. 3d-e). Note however that the overall slope is slightly different for high temperature and low temperature limits.

\textbf{Thermal and thermoelectric response.} We further report that the thermoelectric and thermal response also have geometric contributions as well. In particular, the longitudinal thermal conductance is 
\begin{align}
\kappa_{xx} & =   \frac{1}{T} \sum_{nm, \vec k} \mathcal I^{(2)}_{nm} (\vec k) \, \text{Re} \mathfrak G^{nm}_{ij}  (\vec k) 
\nonumber
\\
& 
-\frac{1}{T} \frac{[ \sum_{nm, \vec k} \mathcal I^{(1)}_{nm} (\vec k) \, \text{Re} \mathfrak G^{nm}_{ij}  (\vec k)  ]^2 }{ [\sum_{nm, \vec k} \mathcal I^{(0)}_{nm} (\vec k) \, \text{Re} \mathfrak G^{nm}_{ij}  (\vec k) ] } ,
\label{flat-kappa}
\end{align}
and thermoelectric power is 
\begin{align}
\Theta_{xx} = \frac{\beta}{e} \frac{ \sum_{nm, \vec k} \mathcal I^{(1)}_{nm} (\vec k) \, \text{Re} \mathfrak G^{nm}_{ij}  (\vec k) }{  \sum_{nm, \vec k} \mathcal I^{(0)}_{nm} (\vec k) \, \text{Re} \mathfrak G^{nm}_{ij}  (\vec k)} ,
\label{flat-thermo}
\end{align}
where we have introduced transport tensors
\begin{align}
\mathcal I^{(\alpha)}_{nm}(\vec k)  = -2  \omega^2_{nm,\vec k} \int  \frac{d \omega}{2 \pi }   \omega^\alpha     f'(\omega)  \,   
\text{Im} \G_{n \vec k} (\omega) \text{Im} \G_{ m \vec k} (\omega) . \nonumber
\end{align}
For numerical purposes, we operate with the same flat band model as discussed below Eq. \eqref{Haldane}, containing two flat Chern bands of band gap $\Delta$ and generic level broadening $\delta_1, \delta_2$ (Fig. 3a). The thermopower is further calculated via \eqref{flat-thermo} and the dimensional units ($k_B, \hbar$) are restored. 
We plot the temperature dependence of thermopower for representative parameters parameters $\delta_{1,2} \ll \Delta$ in Fig. 3 (additional plos with different parameters are listed in SM \cite{SM2022}). The first observation is that the thermopower at the fixed temperature $T \sim \Delta$ is nearly independent of $\delta$ (Fig. 2b), and thus presents a robust quantum transport observable (in this regard, see also  \cite{Kruchkov2020} for thermopower in the SYK flat band). This is in contrast to the flat band conductance (Fig. 2), for which we have found $\sigma_{xx} \propto \delta$, and hence being parameter-dependent. The second observation, is that the thermopower is non-monotonic: it starts nearly linear at low T, but develops a large local maximum $\Theta_*$, placed at $T_* \sim \Delta$. The local maximum presents a "giant" thermopower $\sim 50-60$ $\mu$V/K, unexpected for a narrow Bloch band \cite{Paul2022}.  Above $T_* \sim \Delta$, the thermopower drops to significantly lower values.

\begin{figure}[t]
\center
\includegraphics[width = 1 \columnwidth]{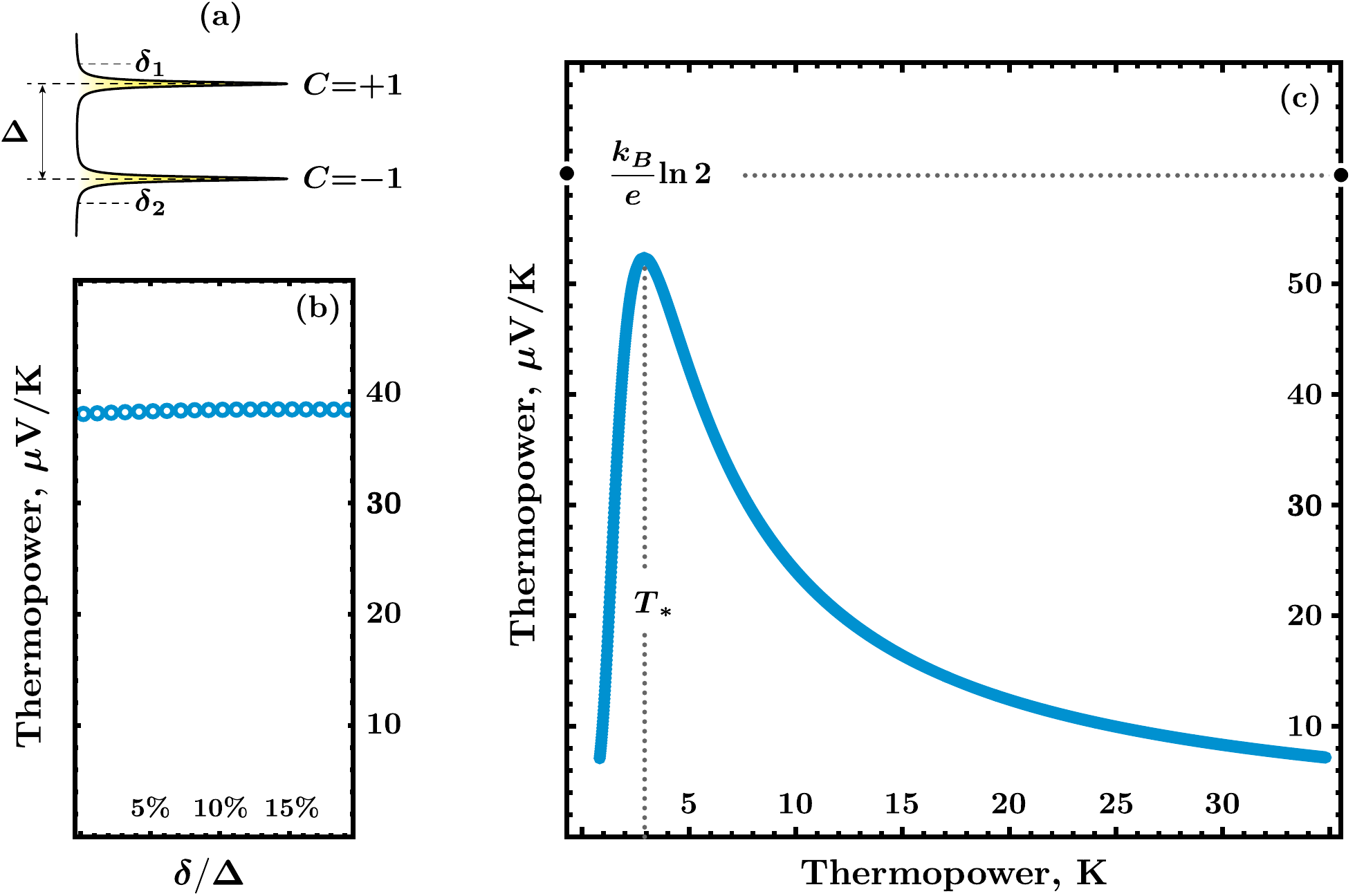}
\caption{\textbf{Anomalous thermoelectricity of topological flat bands}. (a) Schematic of a two-level system with two topological flat bands of $C= \pm 1$. (b) Thermoelectric power $\Theta$ as function of level broadening calculated here for the band gap $\Delta = 0.5$ meV and temperature $T$$=$$\Delta$ ($\approx$ 6 K); here $\delta_{1,2}$$\equiv$$\delta$. (c) Thermoelectric power as a function of temperature, plotted here for $\Delta$$=$$0.5$ meV and $\delta_{1,2}$$=$$0.01$.  Thermopower  develops here a pronounced peak at  $T_*$$\approx$$\Delta /2$, reaching for some parameters the  quantum unit of thermopower $\frac{k_B}{e} \ln 2$. Similar values have been reported for the magic-angle twisted bilayer graphene \cite{Paul2022}, where the system features  narrow topological bands.}
\label{Fig3}
\end{figure}

In the limit when the level broadening is much smaller than the bandgap, the thermopower associated with dispersionless electronic bands with nontrivial quantum geometry  has a bound on local thermopower maximum as  
\begin{align}
\Theta_*  \simeq \frac{k_B}{e} \ln 2 .
\label{theta}
\end{align}
Even a deeper analogy stems here to another case of topological flat bands: in Landau levels \cite{Girvin1982, Jonson1984} the value of  $\frac{k_B}{e} \ln 2 \approx 60 \, \mu \text{V}/\text{K}$ sets the quantum unit for thermopower \footnote{We thank Philip Kim for pointing out  this observation.}. 
As thermopower reflects entropy per quasiparticle,   the quantum unit $\frac{k_B}{e} \ln 2$  reflects fermionic entropy of $k_B \ln 2$ per electron \cite{Hartman2018,Yang2009}, even that there is formally  no Planck constant in Eq. \eqref{theta}.

\textbf{Discussion}. In the absence of magnetic fields, thermopower in graphene-based systems remains  relatively low ($\ll k_B/e$) unless thermally excited to room temperatures \cite{Zuev2009,Nam2010}. In this regard, observation of "giant" thermopower $\sim k_B/e$ in the flat bands of magic-angle twisted bilayer graphene \cite{Paul2022} was unexpected, not captured by conventional theory, and hence attributed to various interaction effects. However, we remark that two slightly gapped flat bands in TBG at the magic angle can be approximated with Landau level wave functions \cite{TKV2019}. Therefore the estimates drawn in this paper for flat band thermopower should qualitatively hold for TBG as well. In particular, the observed large values of thermopower in experiments \cite{Paul2022} are consistent with flat bands thermopower  $\frac{k_B}{e} \ln 2$ (and taking into account the number of flat-band electrons per moiré cell). Moreover, the temperature dependence of thermopower in TBG   (Fig.3c in \cite{Paul2022})   is consistent with temperature dependence stemming from quantum geometry (Fig.\ref{Fig3}).

It is interesting that the superconductivity in perfectly flat Chern bands  \cite{Peotta2015,Huhtinen2022} takes origin from the nontrivial quantum metrics. In fact, the superfluid weight $D_S$ (defined as a response to external vector potential $\vec A$, $j_i = - (D_S)_{ij} A_j$) has a similar linear response structure through current-current correlators \cite{Huhtinen2022}, with a difference that $\vec k \to 0$, $\omega \to 0$ limits should be taken carefully. In this case, one finds 
\begin{align}
D_{S} \sim \Delta_S \sum_{\vec k} \text{Re} \mathfrak G_{ii} (\vec k),
\label{superfluid}
\end{align}
where $\Delta_S$ is the superfluid gap associated with Bogoliubov-de-Gennes model and its variations \cite{Peotta2015,Huhtinen2022,Wang2020}.  After a relevant analysis, the result \eqref{superfluid} has been applied to twisted bilayer graphene, see Refs.\cite{Hazra2019,Hu2019,Julku2020, Xie2020, Guan2022} and Ref.\cite{Tian2021}, where the quantum geometric contribution to superfluid weight, and hence the $T_{\text{BKT}} \sim \frac{\hbar^2}{e^2} D_S (T$$=$$0)$ \cite{Nelson1977}, is argued to be significant (typically, this mechanism gives $T_{\text{BKT}} \sim 1$ K in TBG). As a final remark, there is  an observation that the slope of linear-in-T resistivity in strange metals happens to correlate with the values of superfluid weight (see e.g. Fig. 4 in the recent review by Phillips and colleagues \cite{Phillips2022}); in the quantum transport formalism in flat bands,  this feature may take place from nontrivial $\sum_{\vec k} \text{Re} \mathfrak G_{ij} (\vec k)$.

\

\textbf{Conclusion.} In this work, we have extended the quantum transport formalism for the dispersionless electronic bands. Since the Fermi velocities nullify in  the perfectly flat bands, the conventional  contribution to transport coefficients vanishes. Nevertheless, the flat bands develop an unconventional contribution to transport \eqref{flat-sigma}, which sources from nontrivial   quantum metrics of these electronic states  \eqref{metric}. This contribution leads to enhanced thermoelectricity \eqref{theta} and superconductivity \eqref{superfluid} in the flat band. The origin of these quantum phenomena roots to the Wannier orbital quantization of the topological flat band states: The Wannier orbital cross-section is quantized in Chern numbers, $r_0^2 = C a^2$. This leads to the spatial overlap of real-space electronic orbitals (Fig.1),   and allows quantum tunneling under external fields, as witnessed in the purely quantum contributions to electric conductivity, thermal conductivity, thermoelectric response, and the superfluid weight. 
It will be interesting to extend this formalism towards the topological flat-band systems with SYK interactions \cite{Brzezinska2022}.

\

\textit{Acknowedgments.} The author thanks Subir Sachdev, Philip Kim,  and Bertrand Halperin for fruitful discussions. This work was financed by the Branco Weiss Society in Science, ETH Zurich, through the grant on flat bands, strong interactions, and the SYK physics.

\bibliography{Refs}

\end{document}